\documentstyle[psfig,conf_iap,amssymb]{article}

\newcommand{\be}{\begin{equation}}
\newcommand{\ee}{\end{equation}}
\newcommand{\bref}[1]{(\ref{#1})}
\newcommand{\zmax}{z_{\mathrm{max}}}

\newcommand{\GB}{{\mathcal{L}}_{\mathrm{GB}}}
\newcommand{\LB}{{\mathcal{L}}_{\mathrm{B}}}
\newcommand{\MP}{M_{\mathrm{Pl}}}

\begin{document}
\heading{%
Improving the `self-tuning' mechanism with a Gauss-Bonnet
term
} 
\par\medskip\noindent
\author{%
Stephen C. Davis$^1$
}
\address{%
LPT, Universit\'e de Paris-Sud, B\^at. 210, 91405 Orsay CEDEX, France
}

\begin{abstract}
The effects of higher order gravity terms
on a dilatonic brane world model are discussed~\cite{us}. For a
single positive tension flat
3-brane, and one infinite extra dimension, we present a particular class of
solutions with finite 4-dimensional Planck scale and no naked
singularities. A `self-tuning' mechanism for relaxing the cosmological
constant on the brane, without a drastic fine tuning of parameters, is
discussed in this context.
\end{abstract}

\section{Dilatonic brane worlds}

There has been a great deal of interest in `brane world' models, in
which matter
and fundamental gauge interactions are localized on a four-dimensional
spacetime surface or 3-brane, while gravity is free to propagate in the
higher dimensional bulk spacetime~\cite{BW}. In such models, a fine
tuned relation between the bulk curvature and the brane tension 
has to be specified in order to switch off the effective cosmological
constant on the brane. Without any specific dynamical
mechanism to justify it, this fine-tuning may be seen as a new version of the 
cosmological constant problem in this context. 

This was the starting point for a series of efforts aimed at resolving
this fine tuning problem by means of a dynamical mechanism, called
`self-tuning'~\cite{selftune}. 
A static scalar field, $\phi$, which loosely
models the dilaton and moduli fields of string theory, is added to the
bulk. The extra degree of freedom is then used to ensure the
existence of a solution of the dynamical equations with a
zero effective brane cosmological constant, whatever the value of the
brane tension. Typically a conformally flat solution of the form
\be
ds^2 = e^{2A(z)} \eta_{\mu\nu} dx^\mu dx^\nu + dz^2 \ ,
\label{metric}
\ee
with
\be
A(z) \propto  +\ln\left(1+ \frac{|z|}{z_*} \right) \ , \ \ 
\phi(z) \propto \pm \ln\left(1+\frac{|z|}{z_*} \right) \ ,
\label{stsol1}
\ee
is used (we have assumed a $Z_2$-symmetric bulk). Significantly, the
constant $z_*$ is undetermined by the bulk field equations. The boundary
conditions relate $z_*$ to the brane tension, $T$. Since $z_*$ is
arbitrary, these can be satisfied for any value of $T$, and so the brane
tension does not need to be fine tuned.

There are several problems with this mechanism. For example, why should
one value of $z_*$ be favoured over another, and is the solution stable?
A dynamical analysis of the system is required to address these issues. 

However, the model has a far more serious flaw. By integrating over the
fifth dimension we obtain an effective four dimensional theory on the
brane. The effective Planck mass is then
\be
\MP^2 =  M_s^3 \int_0^{\zmax} dz \, e^{2A} \ ,
\label{MP}
\ee
where where $\zmax$ is the maximum value of $z$, so $\zmax = \infty$ if
$z_* > 0$, and $\zmax=|z_*|$ if $z_* <0$. For $z_*>0$ it is obvious that
$\MP$ is never finite. Alternatively if we choose $z_*<0$, $\MP$ is
finite, but the curvature diverges as $z \to \pm z_*$. Thus no solution of
the form \bref{stsol1} is acceptable. A similar problem occurs in a wide
range of conformally flat dilatonic brane world
models~\cite{sing}.

\section{A higher order gravity tensor}

In four dimensions, the vacuum field equations for gravity are taken to be
$G_{ab} + \Lambda g_{ab} = 0$
since this is the most general tensor which 
(a) is symmetric,
(b) depends only on the metric and its first two derivatives,
(c) is divergence free, and
(d) is linear in second derivatives of the metric.

But in five dimensions there is another possibility. Variation of an
action containing the Gauss-Bonnet term,
\be
\GB=R^2 - 4 R_{ab}R^{ab} + R^{abcd} R_{abcd} \ ,
\label{GB}
\ee
gives the second order Lovelock tensor~\cite{lovelock}
\be
H_{ab} = \left( R R_{ab} - 2 R_{ac}R^c{}_b
-2 R^{c d}  R_{acbd} + R_a{}^{c d e} R_{b c d e}\right)
-\frac{1}{4} g_{ab} \GB \ ,
\ee
which also satisfies the above four conditions.

Thus, in the absence of any evidence to the contrary, we should take the
five dimensional vacuum field equations to be 
$G_{ab} + 2\alpha H_{ab} + \Lambda g_{ab} = 0$, where $\alpha$ and
$\Lambda$ are constants.

A further motivation for higher order curvature terms is that they also appear
in the low energy effective field equations arising from most string
theories~\cite{string}. Since brane worlds are motivated by string
theories, it is particularly natural to include the extra terms in the
five-dimensional field equations. In this case we expect 
$\alpha \sim \Lambda^{-1} \sim M_s^{-2}$. 

We will consider a string theory inspired model with the bulk action
\be
S_5 = \frac{M_s^3}{2} \int d^5x \sqrt{-g}  
\left\{ R - \frac{4}{3} (\nabla \phi)^2 
+  \alpha e^{-4\phi/3}\left[ \GB
+ \frac{16}{27} (\nabla \phi)^4 \right] - 2\Lambda e^{4\phi/3} \right\}
\label{S5}
\ee
The corresponding action in the string
frame, which is related to \bref{S5} by a conformal transformation, is
\be
S_{\mathrm{string}} = \frac{M_s^3}{2}\int d^5x  
\sqrt{-\bar g}  e^{-2\phi} \left\{ \bar R +4(\nabla \phi)^2+
\alpha  \left[ \bar \GB + \cdots \right] 
- \Lambda \right\}  \ .
\ee
Note that the $R$, $\GB$, and $\Lambda$ terms have the same couplings to
$\phi$ in the string frame. However, since they have different
dependencies on $g_{ab}$, this is no longer true in the Einstein frame.

The brane contribution to the action, including boundary
terms~\cite{bound,JC} is
\be
S_4 = - M_s^3 \int d^4x \sqrt{-h} \, 
\left\{ [K] + \alpha e^{-4\phi/3}[\LB] +T e^{2\phi/3} \right\} \ ,
\label{S4}
\ee
where
\be
\LB = 2 K K_{ac}K^{ac} - \frac{4}{3} K_{ac}K^{cb}K^a{}_b -\frac{2}{3}K^3 
- 4 G^{(4)}_{ab}K^{ab} \ ,
\ee
$h_{ab}$ is the induced metric on the brane, and $K_{ab}$ is the
extrinsic curvature.

\section{A non-singular solution}

Variation of generalised boundary term~\bref{S4} gives the junction
conditions on brane~\cite{JC}. Note that the resulting expression has
no dependence on the thickness of the brane. This would not be the case
for any other second order combination of $R_{abcd}$.

The bulk field equations are rather complicated, but for a conformally
flat solution~\bref{metric} with the logarithmic
ansatz~\cite{us,greek}
\be
A(z) = x \ln\left(1+ \frac{|z|}{z_*} \right) \ , \ \ 
\phi(z) = \phi_0 -  \frac{3}{2}\ln\left(1+\frac{|z|}{z_*} \right) \ ,
\ee
they all reduce to algebraic equations
\be
\frac{2\alpha}{z_*^2}  = \frac{1-x}{1 - 6x^3  - 4x^2} e^{4 \phi_0/3} > 0 
\ee
\be
\Lambda \alpha =-\frac{3(-40x^5 + 24x^4 - 52x^3+16x^2+3x-1)(1-x)}
{8(1 - 6x^3  - 4x^2)^2} \ .
\label{Leq}
\ee

The expression for the effective four dimensional Planck mass now
includes additional $\alpha$ dependent corrections, but it is
qualitatively similar to expression~\bref{MP}. To obtain a finite
$\MP$ when  $z_*>0$ (i.e.\ no singularities) we need to find solutions
with $x<-1/2$.

\begin{figure}
\centerline{\psfig{file=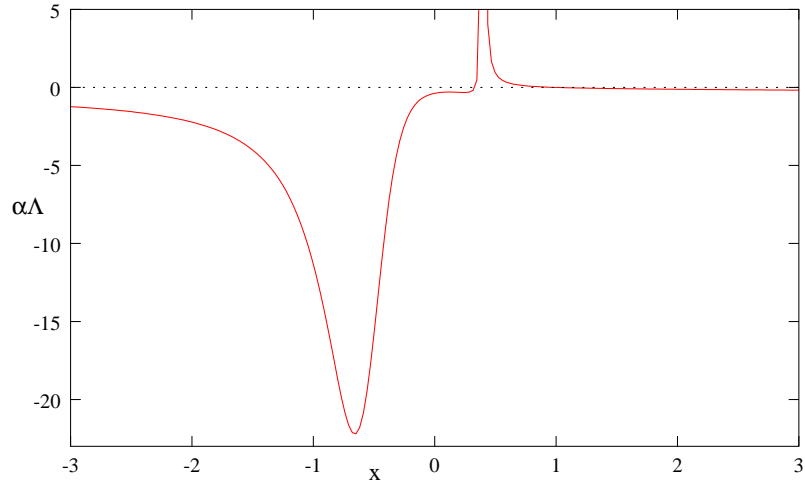,height=2.5in}}
\caption{Variation of $\alpha \Lambda$ with respect to $x$.}
\label{fig}
\end{figure}

Using figure~\ref{fig}, which is a plot of \bref{Leq}, we see that
non-singular solutions with localised gravity exist
if $-22.2 \lesssim \Lambda \alpha< -5/12$~\cite{us}. We expect $\Lambda \alpha
\sim -1$, so this is quite natural. If we set $\alpha =0$, we see that
the only solution is $x=1$, which has infinite $\MP$. Thus
the Gauss-Bonnet term has not removed the singularity
from the $\alpha=0$ solutions, but has instead produced a new branch of
solutions. 

Using the junction conditions we also obtain an algebraic expression
for the brane tension
\be
T = 
\frac{(-x)(3-12x -2x^2-16 x^3)\sqrt{1-x}}{\sqrt{2\alpha}(1-4x^2-6x^3)^{3/2}} 
\, \mathrm{sgn}(z_*) \ .
\label{Teq}
\ee
Thus $T>0$ for $x<0$ when $z_*>0$, so our non-singular, finite $\MP$
solutions do work for models with positive tension branes.
Unfortunately these solutions are not suitable for the `self-tuning'
mechanism, since the value of $T$ is uniquely determined by the value of
$\Lambda$, and so the solution requires fine-tuning after all. It may be
that we need to use a different potential, or it could be that the
mechanism has some flaw which was previously obscured by the
singularity. Further work is required to determine the precise nature of
the problem.

\acknowledgements{I wish to thank Pierre Bin\'etruy, Christos Charmousis, 
Jean-Fran\c{c}ois Dufaux and Jihad Mourad for many useful discussions. I am
also grateful to the conference organisers for giving me the opportunity
to speak. This work was supported by EU network HPRN--CT--2000--00152.}

\vfill
\end{document}